\documentclass[vecphys]{svmult}

\usepackage{makeidx}         
\usepackage{graphicx}        
\usepackage{multicol}        
\usepackage{cite}            
\usepackage[bottom]{footmisc}
\usepackage{amssymb}
\usepackage{amsbsy}

\newcommand{\bkp}{\boldsymbol{\kappa}}

\makeindex             

\begin{document}

\title{Low-Frequency Quantum Oscillations\\
due to Strong Electron Correlations}
\titlerunning{Low-Frequency Quantum Oscillations}
\author{A.~Sherman}
\institute{Institute of Physics, University of Tartu, Riia 142,
51014 Tartu, Estonia\\
\texttt{alexei@fi.tartu.ee}}

\maketitle

\begin{abstract}
The normal-state energy spectrum of the two-dimensional $t$-$J$ model in a homogeneous perpendicular magnetic field is investigated. The density of states at the Fermi level as a function of the inverse magnetic field $\frac{1}{B}$ reveals oscillations in the range of hole concentrations $0.08<x<0.18$. The oscillations have both high- and low-frequency components. The former components are connected with large Fermi surfaces, while the latter with van Hove singularities in the Landau subbands, which traverse the Fermi level with changing $B$. The singularities are related to bending the Landau subbands due to strong electron correlations. Frequencies of these components are of the same order of magnitude as quantum oscillation frequencies observed in underdoped cuprates.
\end{abstract}

\section{Introduction}
Theoretical investigations of systems with strong electron correlations in the magnetic field is of interest in connection with the observation of quantum oscillations in the mixed state of underdoped cuprates \cite{Doiron,Bangura,Yelland}. Based on the Onsager-Lifshitz-Kosevich theory for metals \cite{Shoenberg} the measured decreased quantum oscillation frequencies were interpreted as a manifestation of small Fermi surface pockets \cite{Sebastian12}. If one anticipate that the normal state in the field coincides with the zero-field state, this interpretation seems to contradict numerous photoemission experiments. To explain the appearance of these small pockets proposals for various states with broken translational symmetry were suggested \cite{Millis,Chen,Galitski}. Other theories relate the decreased quantum oscillation frequency to superconducting fluctuations \cite{Melikyan,Pereg} or use phenomenology of the marginal Fermi liquid \cite{Varma}.

Crystals, in which the decreased quantum oscillation frequencies were observed, belong to the underdoped region of the cuprate phase diagram, and, therefore, they are characterized by strong electron correlations. The behavior of models of strong correlations in magnetic fields is poorly known. In this work we consider the two-dimensional (2D) $t$-$J$ model of Cu-O planes. The used approach allows us to take proper account of strong electron correlations and to consider large enough clusters in moderate magnetic fields. An approximation is exploited, in which normal-state holes are characterized by large zero-field Fermi surfaces for hole concentrations $x\gtrsim 0.06$. We found that in the range $0.08<x<0.18$ the density of hole states at the Fermi level oscillates with the inverse magnetic field $\frac{1}{B}$. The oscillations have components with frequencies differing by an order of magnitude. The high-frequency components are connected with the large Fermi surfaces, while the low-frequency components are related to van Hove singularities in the Landau subbands, which traverse the Fermi level with changing $B$. These van Hove singularities are linked with bending the Landau subbands due to strong correlations. Frequencies of slow components are of the same order of magnitude as quantum oscillation frequencies observed in underdoped cuprates. Hence, the $t$-$J$ model supplemented with a mechanism, which smears out high-frequency oscillations, is able to interpret low frequencies of quantum oscillations observed in the experiments.

\section{Main Formulas}
The Hamiltonian of the 2D $t$-$J$ model in the perpendicular magnetic field reads
\begin{equation}\label{hamiltonian}
H=\sum_{\bf ll'\sigma}t_{\bf ll'}\exp\left(i\frac{e}{\hbar}\int_{\bf l}^{\bf l'}\!\!\!{\bf A}({\bf r})\,d{\bf r}\right)a^\dagger_{\bf l\sigma}a_{\bf l'\sigma}
+\frac{1}{2}\sum_{\bf ll'}J_{\bf ll'}\left(s^z_{\bf l}s^z_{\bf l'}+s^+_{\bf l}s^-_{\bf l'}\right),
\end{equation}
where 2D vectors ${\bf l}$ and ${\bf l'}$ label sites of a square plane lattice, $\sigma=\pm 1$ is the projection of the hole spin, $a^\dagger_{\bf l\sigma}=|{\bf l}0\rangle\langle{\bf l}\sigma|$ and $a_{\bf l\sigma}=|{\bf l}\sigma\rangle\langle{\bf l}0|$ are hole creation and annihilation operators with the empty $|{\bf l}0\rangle$ and singly occupied $|{\bf l}\sigma\rangle$ site states. These three states form the complete set of hole states for the site {\bf l} in the $t$-$J$ model. The hole kinetic energy $H_k$ contains the hopping matrix element $t_{\bf ll'}$ and the Peierls phase factor \cite{Peierls} with the vector potential ${\bf A}({\bf r})$. The exchange term includes the exchange constant $J_{\bf ll'}$ and the spin-$\frac{1}{2}$ operators $s^z_{\bf l}=\frac{1}{2}\sum_\sigma \sigma|{\bf l}\sigma\rangle\langle{\bf l}\sigma|$ and $s^\pm_{\bf l}=|{\bf l},\pm 1\rangle\langle{\bf l},\mp 1|$ of localized spins. The Zeeman term is omitted, since for the considered fields and exchange constants in cuprates it is two orders of magnitude smaller than the exchange term.

In the following consideration we suppose that only nearest neighbor hopping and exchange constants are nonzero, $t_{\bf ll'}=t\sum_{\bf a}\delta_{\bf l,l'+a}$, $J_{\bf ll'}=J\sum_{\bf a}\delta_{\bf l,l'+a},$ where ${\bf a}$ are four vectors connecting nearest neighbor sites. In the Landau gauge ${\bf A(l)}=-Bl_y{\bf x}$, where $l_y$ is the $y$ component of the site vector ${\bf l}$ and ${\bf x}$ is the unit vector along the $x$ axis. Hence the exponential in the kinetic term of the Hamiltonian can be written as
\begin{equation}\label{pphase}
{\rm e}^{i\bkp_{\bf a}{\bf l}},\quad \bkp_{\bf a}=-\frac{e}{\hbar}Ba_x{\bf y}.
\end{equation}

We shall restrict our consideration to the fields satisfying the condition
\begin{equation}\label{condition}
\frac{e}{\hbar}Ba^2=2\pi\frac{n'}{n},
\end{equation}
where $a=|{\bf a}|$, $n$ and $n'<n$ are integers with no common factor. In this case the kinetic term of the Hamiltonian defines its translation properties -- $H_k$ is invariant with respect to translations by the lattice period along the $x$ axis and by $n$ lattice periods along the $y$ axis. To retain this symmetry we apply the periodic Born-von Karman boundary conditions to the sample with $N_x$ sites along the $x$ axis and $nN_y$ sites along the $y$ axis. The boundary conditions define the set of allowed wave vectors with components $K_x=\frac{2\pi}{N_xa}n_x$ and $K_y=\frac{2\pi}{nN_ya}n_y$ with integer $n_x$ and $n_y$. As can be seen from (\ref{pphase}) and (\ref{condition}), the momenta $\bkp_{\bf a}$ coincide with some wave vectors in this net. This allows us to perform the usual Fourier transformation, $a_{\bf l\sigma}=N^{-1/2}\sum_{\bf K}{\rm e}^{-i{\bf Kl}}a_{\bf K\sigma}$, $N=nN_xN_y$, and obtain for $H_k$
\begin{equation}\label{kterm}
H_k=t\sum_{\bf Ka\sigma}{\rm e}^{i{\bf Ka}}a^\dagger_{{\bf K}-\bkp_{\bf a},\sigma}a_{\bf K\sigma}.
\end{equation}

It is convenient to split the Brillouin zone into $n$ stripes of the width $\frac{2\pi}{na}$, which are oriented parallel to the $x$ axis. If we select the lowest stripe with $-\frac{\pi}{a}<K_y\leq -\frac{\pi}{a} +\frac{2\pi}{na}$, and denote wave vectors in it as ${\bf k}$,  momenta in the entire Brillouin zone can be described as ${\bf k}+j\bkp$. Here $0\leq j \leq n-1$ and $\bkp=\frac{2\pi}{na}{\bf y}$. In these notations the kinetic energy acquires the form $H_k=\sum_{\bf k\sigma}{\bf A^\dagger_{k\sigma}h_k A_{k\sigma}}$, where the summation over ${\bf k}$ is performed over the selected stripe, ${\bf A^\dagger_{\bf k\sigma}}=\left(a^\dagger_{\bf k\sigma},a^\dagger_{\bf k+\bkp,\sigma},\ldots a^\dagger_{{\bf k}+(n-1)\bkp,\sigma}\right)$ and the matrix ${\bf h_k}$ has the following elements:
\begin{equation}\label{hmatr}
h_{{\bf k}jj'}=\left\{
\begin{array}{ll}
2\cos\left(k_y a+\frac{2j\pi}{n}\right), & j=j', \\
{\rm e}^{-ik_xa}, & j=j'+n', \\
{\rm e}^{ik_xa}, & j=j'-n', \\
0, & {\rm in\ other\ cases.}
\end{array}
\right.
\end{equation}
In this equation the matrix indices $j$ and $j'$ are determined modulo $n$.

The Hermitian matrix (\ref{hmatr}) can be diagonalized by the unitary transformation ${\bf U_k}$. Since the kinetic energy defines symmetry properties of the total Hamiltonian (\ref{hamiltonian}), the operators
\begin{equation}\label{alpha}
\alpha_{{\bf k}m\sigma}=\sum_{j=0}^{n-1}U^*_{{\bf k}jm}a_{{\bf k}+j\bkp,\sigma}, \quad 0\leq m\leq n-1,
\end{equation}
form a basis for a representation of the symmetry group of the Hamiltonian. Besides, in the absence of correlations states created by $\alpha_{{\bf k}m\sigma}$ are eigenstates of the Hamiltonian. Therefore, Green's function constructed from these operators,
\begin{equation}\label{green}
G({\bf k}m\bar{t})=-i\theta(\bar{t})\left\langle\left\{\alpha^\dagger_{{\bf k}m\sigma}(\bar{t}),\alpha_{{\bf k}m\sigma}\right\}\right\rangle,
\end{equation}
is an appropriate mean for investigating the influence of strong correlations on hole states. In (\ref{green}), the averaging over the grand canonical ensemble and the operator time dependence are determined by the Hamiltonian ${\cal H}=H-\mu \sum_{\bf l}|{\bf l}0\rangle\langle{\bf l}0|$ with the chemical potential $\mu$.

To calculate (\ref{green}) we use the Mori projection operator technique \cite{Mori}. In this approach, the Fourier transform of Green's function (\ref{green}) is represented by the continued fraction
\begin{equation}\label{cfraction}
G({\bf k}m\omega)=\frac{\left\langle\left\{\alpha_{{\bf k}m\sigma}, \alpha^\dagger_{{\bf k}m\sigma}\right\}\right\rangle
}{\omega-E_0-\frac{\textstyle V_0}{\textstyle\omega-E_1-\frac{\textstyle V_1}{\ddots}}},
\end{equation}
where
\begin{eqnarray}
&&\left\langle\left\{\alpha_{{\bf k}m\sigma},\alpha^\dagger_{{\bf k}m\sigma}\right\}\right\rangle=\phi,\label{numerator}\\
&&E_0=\sum_{j=0}^{n-1}\sum_{\bf a}U^*_{{\bf k}jm}U_{{\bf k},j+j_{\bf a},m}
\bigg[{\rm e}^{i({\bf k}+j\bkp){\bf a}}\bigg(t\phi\
+\frac{3tC_1}{2\phi}\bigg)+\frac{tF_1}{\phi}\bigg]\nonumber\\
&&\qquad -\frac{3JF_1}{\phi}\sum_{j=0}^{n-1}|U_{{\bf k}jm}|^2\gamma_{{\bf k}+j\bkp}-\frac{3JC_1}{\phi}-\mu,\label{e0}\\
&&V_0=\sum_{j=0}^{n-1}\sum_{\bf aa'}U^*_{{\bf k}jm}U_{{\bf k},j+j_{\bf a}-j_{\bf a'},m}{\rm e}^{i({\bf k}+j\bkp)({\bf a-a'})-i\bkp_{\bf a}{\bf a'}}\nonumber\\
&&\qquad\times t^2\bigg(\phi^2+\frac{3C_1}{2\phi}+\frac{3C_{\bf a-a'}}{2}\bigg)\nonumber\\
&&\qquad -t(\mu+E_0)\sum_{j=0}^{n-1}\sum_{\bf a}U^*_{{\bf k}jm}U_{{\bf k},j+j_{\bf a},m}
\bigg[{\rm e}^{i({\bf k}+j\bkp){\bf a}}
\bigg(2\phi+\frac{3C_1}{\phi}\bigg)+\frac{2F_1}{\phi}\bigg]\nonumber\\
&&\qquad+\frac{t^2F_1}{\phi}{\rm Re}\sum_{j=0}^{n-1}\sum_{\bf aa'}U^*_{{\bf k}jm}U_{{\bf k},j+j_{\bf a}-j_{\bf a'},m}{\rm e}^{i({\bf k}+j\bkp){\bf a}}\nonumber\\
&&\qquad -\frac{t^2F_1}{\phi}\sum_{j=0}^{n-1}\sum_{\bf a}U^*_{{\bf k}jm}U_{{\bf k},j+2j_{\bf a},m}{\rm e}^{i({\bf k}+j\bkp){\bf a}}
\nonumber\\
&&\qquad +(\mu+E_0)^2+t^2x-\frac{4t^2C_1}{\phi},\label{v0}\\
&&E_1\approx -\mu. \label{e1}
\end{eqnarray}
In the above equations, $\phi=\frac{1+x}{2}$, $j_{\bf a}=-n'\frac{a_x}{a}$,
\begin{equation}\label{x}
x=\frac{1}{N}\sum_{\bf l}\Big\langle|{\bf l}0\rangle\langle{\bf l}0|\Big\rangle
=-\frac{1}{N\pi} \sum_{{\bf k}m}\int_{-\infty}^\infty d\omega\frac{{\rm Im}G({\bf k}m\omega)}{{\rm e}^{\beta\omega}+1}
\end{equation}
is the hole concentration, $\gamma_{\bf k}=\frac{1}{4}\sum_{\bf a}{\rm e}^{i{\bf ka}}$,
\begin{equation}\label{f1}
F_1=\frac{1}{4N}\sum_{\bf la}\left\langle a^\dagger_{\bf l\sigma} a_{\bf l+a,\sigma}\right\rangle
=\frac{1}{N}\sum_{{\bf k}jm}\gamma_{{\bf k}+j\bkp}|U_{{\bf k}jm}|^2 \int_{-\infty}^\infty d\omega\frac{{\rm Im}G({\bf k}m\omega)}{{\rm e}^{\beta\omega}+1},
\end{equation}
$C_1=\frac{1}{4N}\sum_{\bf la}\left\langle s^+_{\bf l}s^-_{\bf l+a} \right\rangle$ and $C_{\bf a-a'}=\frac{1}{N}\sum_{\bf l}\left\langle s^+_{\bf l}s^-_{\bf l+a-a'}\right\rangle$.

In the course of calculations parameters $x$ and $F_1$ were determined self-consistently. Parameters $C_1$ and $C_{\bf a-a'}$, which characterize the spin subsystem and are indirectly influenced by the field, were taken from the results of zero-field calculations \cite{Sherman04}. The quantity $E_1$ is described by a more complex expression than that given by (\ref{e1}). The comparison of results obtained with this expression and with (\ref{e1}) shows their similarity. Therefore, to speed up calculations we used the latter value for $E_1$. Notice that (\ref{hmatr}), (\ref{e0}) and (\ref{v0}) generalize equations of ref.~\cite{Sherman13} to the case $n'\neq 1$, which allows us to vary $\frac{1}{B}$ with a step as small as is wished.

Terminating calculations of the continued fraction (\ref{cfraction}) with the term $E_1$ Green's function (\ref{green}) is approximated by the expression with two poles for every wave vector ${\bf k}$ and subband index $m$. This procedure gives the first Mori correction to the uncorrelated solution described by the operator $\alpha_{{\bf k}m\sigma}$. The correction takes into account electron correlations. In view of rapid convergence of utmost poles of a finite continued fraction with the number of its terms, this procedure is supposed to give a qualitatively correct description for these poles. This supposition is supported by the results obtained with this approximation for $B=0$ \cite{Sherman04}. In this case the pole with the lower energy corresponds to the spin-polaron band, which gives the most intensive maximum in the hole spectral function. The two-pole approximation provides a satisfactory description of the evolution of this band with doping. For $B\neq 0$ the states near the Fermi level originate from this band, which gives grounds to use the same approximation for their description.

Notice, however, that the comparatively small number of calculated continued fraction elements does not allow us to describe the pseudogap and Fermi arcs in the zero-field hole spectrum. In the used approximation the $B=0$ Fermi surface is large for $0.06\lesssim x\lesssim 0.18$, having shape of diamonds with rounded corners, which are centered at $\left(\pm\frac{\pi}{a},\pm \frac{\pi}{a}\right)$. In the $t$-$J$ model, at moderate doping the pseudogap and Fermi arcs reflect the dispersions of the spin-polaron band with the bandwidth $\sim J$ and the spectrally less intensive band with the bandwidth $\sim t\gg J$ \cite{Sherman97}. This latter band has the mentioned large Fermi surface. Therefore, we suppose that the used approximation gives at least qualitatively a correct picture of the energy spectrum at $B\neq 0$.

\section{Results and Discussion}
\begin{figure}[t]
\centering
\includegraphics[width=.7\textwidth]{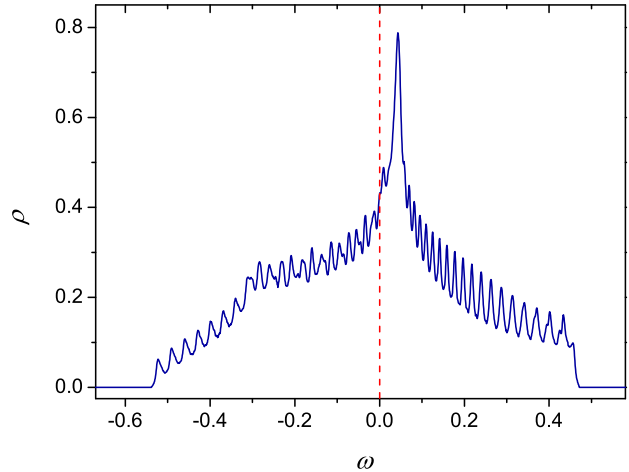}
\caption{The density of hole states in the case $n=47$, $n'=1$ and $x=0.14$. The Fermi level is shown by the red dashed line.} \label{Fig1}
\end{figure}

For $B\neq 0$ the density of the hole states (DOS),
\begin{equation}\label{dos}
\rho(\omega)=-\frac{1}{N\pi}\sum_{{\bf k}m}{\rm Im}G({\bf k}m\omega),
\end{equation}
as a function of frequency and magnetic field reveals oscillations near the Fermi level. An example of these oscillations is shown in Fig.~\ref{Fig1}.
Here and hereafter we set $t$ and $a$ as units of energy and length, respectively. In the calculations we set $J/t=0.2$ and the temperature $T=0$. The oscillations are observed in the concentration range $0.08<x<0.18$. Outside of this range the Fermi level falls on strong maxima of the DOS, where the oscillations are lost against the background.

\begin{figure}[t]
\centering
\includegraphics[width=.6\textwidth]{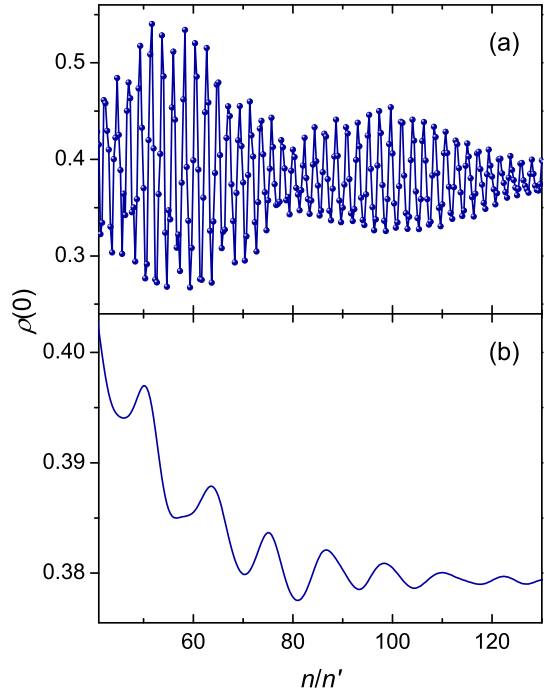}
\caption{(a) The density of hole states at the Fermi level as a function of $\frac{n}{n'}=\frac{2\pi\hbar}{ea^2}\frac{1}{B}$ for $n'=3$ and $x=0.14$. Calculated values are shown by symbols, connecting lines are a guide to the eye. (b) The same data smoothed by the FFT filter method.} \label{Fig2}
\end{figure}
As mentioned above, in the considered range of $x$ the zero-field Fermi surfaces are large. The calculated period of oscillations of $\rho(0)$ as a function of $\frac{n}{n'}=\frac{2\pi\hbar}{ea^2}\frac{1}{B}$ conforms with such Fermi surfaces -- for the case $x=0.14$ the period is equal to $2-3$ [see Fig.~\ref{Fig2}(a)] in the units of $\frac{n}{n'}$, which corresponds to the Onsager frequency $F\sim 10$~kT for $a=4$~\AA (an approximate distance between copper sites in Cu-O planes). However, along with these high-frequency oscillations a modulation with a period which is larger by an order of magnitude is also observed. To reveal the respective low-frequency oscillations the data were smoothed with the FFT filter method, which allows one to suppress high-frequency components. The result is shown in Fig.~\ref{Fig2}(b). The frequency of these oscillations is of the order of 1~kT, which is comparable to the dominant frequency of quantum oscillations in underdoped YBa$_2$Cu$_3$O$_{6+x}$ \cite{Sebastian12}. Thus, supplemented with a mechanism, which smears out high-frequency oscillations, the $t$-$J$ model is able to explain the low-frequency quantum oscillations observed in cuprate perovskites. As candidates for such a mechanism finite lifetimes of hole states, hole spins, field and crystal inhomogeneities \cite{Shoenberg} can be mentioned. For other considered hole concentrations in the range $0.08<x<0.18$ similar pictures of high- and low-frequency oscillations in the DOS can be revealed.

\begin{figure}
\centering
\includegraphics[width=.65\textwidth]{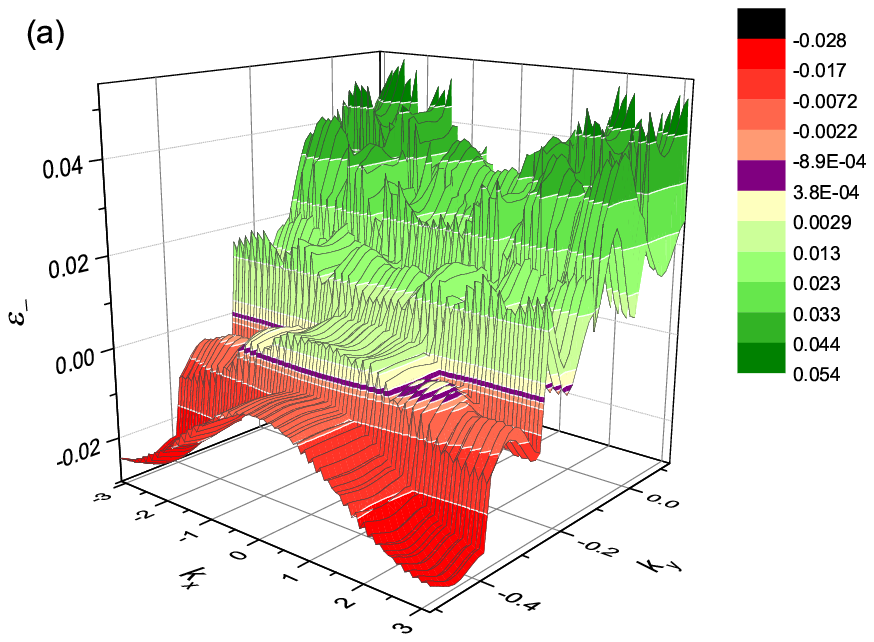}
\includegraphics[width=.65\textwidth]{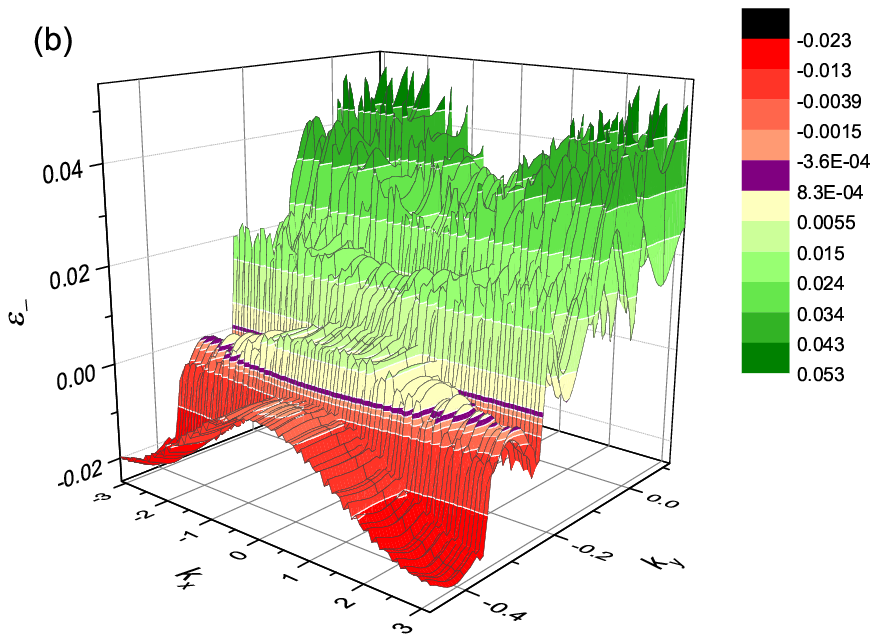}
\includegraphics[width=.65\textwidth]{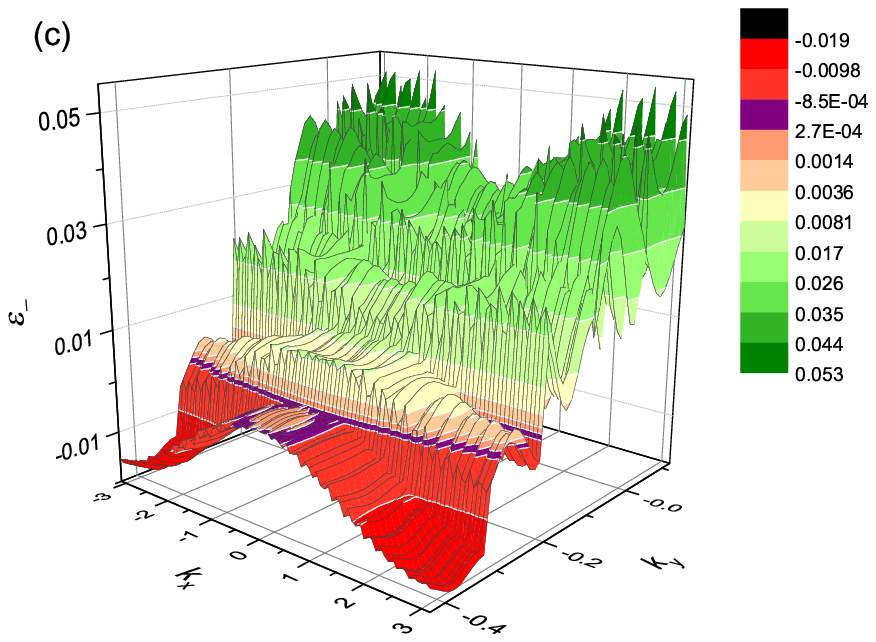}
\caption{The dispersion of hole states near the Fermi level for $n=48$ (a), 52 (b) and 56 (c). $x=0.14$ and $n'=1$. The Fermi level is shown by the purple (dark) contour.}\label{Fig3}
\end{figure}
What is the reason for the appearance of the low-frequency modulation in Fig.~\ref{Fig2}(a) and oscillations in Fig.~\ref{Fig2}(b)? In the uncorrelated case carrier energies form a stair of nearly {\bf k}-independent Landau subbands (in which energies experience weak oscillations when $k_x$ varies from $-\pi$ to $\pi$ at a fixed $k_y$ in the Landau gauge). The traverse of the Fermi level through these subbands leads to oscillations in the DOS with a nearly constant amplitude. Strong correlations lead to a bend of the subbands along the $x$ direction. This bending is seen in Fig.~\ref{Fig3}. In this figure, only the subbands near the Fermi level are shown for the case $n'=1$. The subband $m$ is plotted in the stripe $-\pi+\frac{2\pi}{n}m<K_y\leq-\pi+\frac{2\pi}{n}(m+1)$, $-\pi< K_x\leq\pi$ of the entire Brillouin zone. As a result the hole dispersion looks like stairs with steps in the $y$ direction. These stairs ascend from smaller hole energies to larger ones with $K_y$ moving from $-\pi$ to $\pi$. The small-scale oscillations along the $x$ direction have the same origin as in the uncorrelated case. Bends of the subbands lead to the appearance of van Hove singularities, and these singularities supplement the DOS oscillations with the amplitude modulation seen in Fig.~\ref{Fig2}(a). Panels of Fig.~\ref{Fig3} demonstrate cases when the Fermi level falls onto a van Hove singularity [panel (c)], is halfway between two singularities [panel (a)] and is located in some intermediate position [panel (b)]. Thus, the low-frequency modulation and oscillations in Fig.~\ref{Fig2} are connected with the traverse of the Fermi level through the sequence of van Hove singularities in the Landau subbands.

\section{Concluding Remarks}
In summary, we have considered the two-dimensional $t$-$J$ model of Cu-O planes of cuprate perovskites under the conditions of strong electron correlations: $t\gg J$ and small hole concentrations $x$. The two-dimensional crystal is placed in a magnetic field, which is perpendicular to the crystal plane. Using the Mori projection operator technique we have calculated the Landau subbands in the case when only the nearest neighbor hopping constant is nonzero and for the magnetic field induction satisfying the condition $B=\frac{2\pi\hbar}{ea^2}\frac{n'}{n}$, where $a$ is the lattice spacing, $n$ and $n'<n$ are integers with no common factor. In the range of hole concentrations $0.08<x<0.18$ the density of hole states at the Fermi level $\rho(0)$ shows oscillations as a function of $\frac{1}{B}$. For somewhat smaller and larger hole concentrations the Fermi level falls onto strong maxima of the density of states, which hide the oscillations. The oscillations have high- and low-frequency components. Frequencies of the former components conform with large zero-field Fermi surfaces, which are inherent in the model for these $x$ in the used approximation. The components with an order of magnitude smaller frequencies are connected with the bending of the Landau subbands near the Fermi level. This bending is a result of strong electron correlations. The bending leads to the appearance of van Hove singularities in the Landau subbands, which traverse the Fermi level with changing $B$. Onsager frequencies of these components $F\sim 1$~kT are of the same order of magnitude as dominant quantum oscillation frequencies observed in underdoped cuprate perovskites. Being supplemented with a mechanism, which smears out the high-frequency components, the calculated oscillations of $\rho(0)$ become similar to quantum oscillations in these experiments.

\section*{Acknowledgements}
This work was supported by the European Regional Development Fund (project TK114) and by the Estonian Scientific Foundation (grant ETF9371).


\printindex
\end{document}